\documentclass[floats,floatfix,showpacs,amssymb,prd,superscriptaddress,twocolumn,aps]{revtex4-1}

\usepackage{graphicx}  % needed for figures
\usepackage{dcolumn}   % needed for some tables
\usepackage{bm}        % for math
\usepackage{amssymb}   % for math

\usepackage{lipsum}

\usepackage{epsf}
\usepackage{epstopdf}

\usepackage{graphicx}

\usepackage[linktocpage]{hyperref}
\usepackage[caption=false]{subfig}
\usepackage[usenames]{color}

\newcommand{\insertplot}[5]{\begin{figure}
 \hfill\hbox to 0.05in{\vbox to #5in{\vfill
 \inputplot{#1}{#4}{#5}}\hfill}
 \hfill\vspace{-.1in}
 \caption{#2}\label{#3}
 \end{figure}}
 \newcommand{\inputplot}[3]{
 \special{ps: plotfile #1}
}
\newcounter{fig}

\usepackage{graphicx}

\newcommand{\abs}[1]{\ensuremath{\left\vert#1\right\vert}} % Betragsstriche mit /abs{Argument}

\begin{document}

\title{Axial quasinormal modes of static neutron stars \\ 
in the nonminimal derivative coupling sector of Horndeski gravity: \\
Spectrum and universal relations for realistic equations of state}

\author{Jose Luis Bl\'azquez-Salcedo}
\email{jose.blazquez.salcedo@uni-oldenburg.de}
\affiliation{Institut f\"{u}r Physik, Universit\"{a}t Oldenburg, 26111 Oldenburg, Germany}

\author{Kevin Eickhoff}
\email{kevin.eickhoff@uni-oldenburg.de}
\affiliation{Institut f\"{u}r Physik, Universit\"{a}t Oldenburg, 26111 Oldenburg, Germany}

\date{\today}

\begin{abstract}
	We study axial quasinormal modes of static neutron stars in the nonminimal derivative coupling sector of Horndeski theory. We focus on the fundamental curvature mode, which we analyze for 10 different equations of state with different matter content. A comparison with the results obtained in pure general relativity reveals that, apart from modifying the spectrum of the frequencies and the damping times of the stars, this theory modifies several universal relations between the modes and physical parameters of the stars that are otherwise matter independent.  	
\end{abstract}

\pacs{}
\maketitle

\section{Introduction}

With the direct detections of gravitational waves (GW) by the LIGO-VIRGO Collaboration \cite{ligo1,ligo_2nd,ligo_3rd,ligo_4th,ligo_neutron}, a new era in astronomy has begun, no longer constrained by the electromagnetic channel of observation.
These detections allow us to observe phenomena in the strong-field gravity regime. 
The first four events detected fit very accurately with pairs of astrophysical black hole mergers \cite{ligo1,ligo_2nd,ligo_3rd,ligo_4th}. 
The fifth detection GW170817, however, presents the characteristics of a pair of neutron stars colliding and merging \cite{ligo_neutron}, and the event was also observed as a $\gamma$-ray burst (GRB 170817A).
In this case, the nature of the remnant object is unknown, and it could be a black hole, an unstable massive neutron star (that eventually collapses into a black hole), or a stable neutron star \cite{Abbott:2017dke,Piro:2017zec}. The remnant is expected to radiate GW after the merger during the ringdown phase with some specific frequencies and damping times. 
Although with current detectors the ringdown phase of the last GWs could not be very well resolved, future enhancements in the sensitivity of the experiments will likely allow for the observation of the ringdown frequencies and damping times of similar events.
Hence, it is important to have a good theoretical understanding of the ringdown phase of these objects, so that in the future a detection can be compared with theoretical predictions. The spectrum of the ringdown phase of neutron stars can be studied by using a quasinormal mode (QNM) decomposition \cite{Kokkotas:1999bd,Nollert:1999ji,Ferrari:2007dd}. Because of the presence of matter, the QNM spectrum of a neutron star is different and richer than the one of a black hole, despite the absence of a horizon. The particular frequencies and damping times of the ringdown depend on the matter composition of the star, which is unknown in the inner regions of the star and parametrized by an equation of state (EOS) \cite{Lattimer:2012nd,Haensel:2007yy,Heiselberg:1999mq}. 
Nevertheless, there exist a number of universal relations between the QNM spectrum and physical quantities of the star that are matter independent \cite{Kokkotas:1999mn,Benhar:2004xg,BlazquezSalcedo:2012pd,Blazquez-Salcedo:2013jka}. These are similar in spirit to the ''I-love-Q'' universal relations, that exist between several rotational parameters of the star \cite{Yagi:2013bca}. These universal relations hold approximately for most realistic matter models in general relativity (GR). However, if alternative theories of gravity are considered \cite{Doneva:2017jop,Berti:2015itd}, such relations could in principle present deviations from GR, deviations that could be used together with observations to constrain the alternative theories \cite{tests,insp_binar,Blazquez-Salcedo:2015ets}.
In this paper we consider the case of a particular type of scalar-tensor theory. The most general scalar-tensor theory with only second order derivatives in the field equations is the Horndeski theory \cite{horndeski}, which avoids the appearance of ghosts and maintains global hyperbolicity. The theory has been rediscovered recently as covariant Galileon gravity in four dimensions \cite{Nicolis:2008in,Kobayashi:2011nu}. 
Neutron stars in Horndeski and beyond Horndeski have been studied in  \cite{Maselli:2016gxk,Babichev:2016jom}.
A subclass of these theories known as ''Fab-Four'' \cite{fab4} (consisting of the sum of four actions) possess solutions describing compact objects that could have some astrophysical interest.
The combination of the ''John'' and ''George'' terms is known as the nonminimal derivative coupling sector of Horndeski theory. Therefore the action is
%\begin{widetext}
\begin{eqnarray}
\label{action} 
&S&=\int  d^4x  \sqrt{-g} \bigg[ \kappa (R-2\Lambda)   \\
&& -\frac{1}{2}(\alpha g^{\mu\nu}-\eta G^{\mu\nu}) \nabla_{\mu} \phi \nabla_{\nu} \phi \bigg]
+ S_m, \nonumber
\end{eqnarray}
%\end{widetext}
where $g$ is the metric and $\phi$ is the scalar field, with $R$ the Ricci scalar and $G_{\mu \nu}$ the Einstein tensor. The constants $\alpha$ and $\eta$ are in principle free, and $\kappa = \frac{c^4}{16 \pi G}$.
$S_m$ is the action describing ordinary matter.

Let us note here that several black hole solutions and their properties have been studied in this theory \cite{Rinaldi:2012vy,self_tune,Charmousis:2015aya,2014PhRvD..89f4017M,2014PhRvD..89h4038C,2014PhRvD..89h4042K,2014PhRvD..89h4050A,2014PhRvD..89j4028G,2014JHEP...07..085C,2014PTEP.2014g3E02K,2014JHEP...08..106B,2014GReGr..46.1785M,2015JPhCS.600a2003C,Ogawa:2015pea,Takahashi:2016dnv,Dong:2017toi}. 

Although a post-Newtonian analysis has shown that the theory is heavily constrained by Solar System tests  \cite{Bruneton:2012zk}, 
asymptotically flat neutron stars have been obtained for the 
action (\ref{action}) in the $\Lambda=\alpha=0$ case \cite{scalar,ludovic}. Interestingly, outside the star the metric is essentially described by the GR vacuum solution, which means that these solutions could be astrophysically relevant. Nonetheless, the scalar field couples to the matter content of the star. The configuration possesses a nontrivial scalar field in its interior that modifies the matter distribution and affects the global quantities of the configuration.
Such models of neutron stars have been studied first in the static and spherically symmetric case in \cite{scalar}, and later in the slow rotation limit in \cite{ludovic}.

The purpose of this paper is to calculate the spectrum of axial quasinormal modes of static and spherically symmetric neutron stars obtained in \cite{scalar}. In section II we briefly review the field equations, Ansatz and results for the static and spherically symmetric neutron stars with the new equations of state we consider. In section III we present the differential equations and boundary conditions describing the axial perturbations. In section IV we present the spectrum of quasinormal modes and we compare with the results from GR by calculating deviations in several matter-independent universal relations between scaled parameters.

\section{Neutron stars}
\label{sec2}

From variations of the action (\ref{action}) with $\Lambda=\alpha=0$, we obtain the following set of field equations
\begin{equation}
\label{eq:field_eq}
G_{\mu \nu} - H_{\mu \nu} = \frac{1}{2 \kappa} T_{\mu \nu}, \ \
\nabla_{\mu}J^{\mu} =0, 
\end{equation}
where the tensor $H_{\mu\nu}$ is
\begin{eqnarray}
H_{\mu \nu} &=& \frac{\eta}{2 \kappa} \bigg\lbrace \frac{1}{2} \nabla_{\mu}\phi \nabla_{\nu} \phi R - 2 \nabla_{\lambda} \phi \nabla_{( \mu} \phi R^{\lambda}_{\nu )} \nonumber \\
&-& \nabla^{\lambda} \phi \nabla^{\rho}\phi R_{\mu \lambda \nu \rho} - (\nabla_{\mu}\nabla^{\lambda}\phi)(\nabla_{\nu}\nabla_{\lambda}\phi)\nonumber \\
&+& \frac{1}{2}g_{\mu \nu}(\nabla^{\lambda} \nabla^{\rho}\phi)(\nabla_{\lambda} \nabla_{\rho} \phi) - \frac{1}{2}g_{\mu \nu}(\square \phi)^2 \\
&+& (\nabla_{\mu} \nabla_{\nu}\phi)(\square \phi) + \frac{1}{2}G_{\mu \nu}(\nabla \phi)^2 %
+ g_{\mu \nu} \nabla_{\lambda}\phi \nabla_{\rho}\phi R^{\lambda \rho} \bigg\rbrace
\nonumber 
\end{eqnarray}
and the scalar current $J_{\mu}$ is
\begin{equation}
J^{\mu} =  - \eta G^{\mu \nu} \nabla_{\nu}\phi.
\end{equation}

The energy-momentum tensor $T_{\mu\nu}$ parametrizes the matter inside the star, which we will suppose to be a perfect fluid with energy density $\rho$, pressure $P$, and four-velocity $u$: 
\begin{equation}
T_{\mu \nu} = (\rho + P)u_{\mu}u_{\nu} + P g_{\mu \nu}.
\end{equation}
To construct spherically symmetric neutron stars, we follow \cite{scalar} and use a metric Ansatz of the form
\begin{equation}
\label{metric}
ds^2 = -b(r)dt^2 + \frac{1}{f(r)}dr^2 + r^2 \sin^2(\theta) d\varphi^2 + r^2 d\theta^2,
\end{equation}
where the functions $b(r)$ and $f(r)$ are in principle arbitrary, to be determined by the field equations and boundary conditions.
A compatible Ansatz for the scalar field takes the form \cite{2014JHEP...08..106B}
\begin{equation}
\label{rest_field}
\phi(r,t) = Qt+F(r),
\end{equation}
with $F(r)$ as an arbitrary function of the radial coordinate. 
Note that in principle one can allow for a scalar field with some linear time dependence controlled by the ''clock'' parameter $Q$. This is compatible with the metric (\ref{metric}), since the field equations only depend on $\nabla \phi$, and only the $Q$ parameter enters the field equations. We will keep this parameter free in the calculations of the neutron star models, but in our calculation of the QNMs we will focus on static solutions, setting $Q=0$. 

In vacuum we have $T_{\mu\nu}=0$, which implies that $H_{\mu\nu}=0$. Hence a possible solution outside of the star is simply given by the Schwarzschild metric with a nontrivial scalar field determined by the conservation of the scalar current from equation (\ref{eq:field_eq}). These solutions form a family of asymptotically flat black hole solutions called ''stealth'' configurations and are studied in detail in \cite{2014JHEP...08..106B}.

These solutions describe also the exterior of the neutron stars. However inside the neutron stars $T_{\mu\nu}\neq 0$ and $H_{\mu\nu}\neq 0$. This results in a modification of the Tolman-Oppenheimer-Volkoff (TOV) equations:
\begin{eqnarray}
\label{tov1}
&& P' = -(\rho + P) \frac{b'}{2b}, \ \ \ \
b' = \frac{b(1-f)}{fr}, \\
&& f' = - \frac{3Q^2 \eta f(1-f) + b(6Pfr^2 + \rho r^2(1+f)-4 \kappa(1-f)) }{rb(Pr^2+4 \kappa)- 3Q^2 \eta f r }. \nonumber 
\end{eqnarray}
In addition we have the conservation of the scalar current (\ref{eq:field_eq}), which gives us the equation for the scalar field radial function $F(r)$
\begin{equation}
\label{tov4}
\eta f b (F')^2 = P b r^2 + Q^2 \eta (1-f).
\end{equation}
These modified TOV equations cannot be smoothly connected with the standard GR equations \cite{scalar,ludovic}, and the neutron stars possess necessarily nontrivial scalar hair. 

It is possible to obtain a perturbative solution at the center $r=0$ of the star, given by the following expressions:
\begin{eqnarray}
\label{pert_sol1}
b= b_0 -b_0^2\frac{3P_c+\rho_c}{3(3Q^2\eta-4\kappa b_0)}r^2 + O(r^4), \nonumber\\
f= 1+\frac{2b_0(3P_c+\rho_c)}{3(3Q^2\eta-4\kappa b_0)}r^2 + O(r^4), \nonumber\\
P= P_c + b_0\frac{(P_c+\rho_c)(3P_c+\rho_c)}{6(3Q^2\eta-4\kappa b_0)}r^2 + O(r^4), \nonumber\\
(F')^2 = \left( \frac{P_c}{\eta} -\frac{2(3P_c+\rho_c)Q^2}{3(3Q^2\eta-4\kappa b_0)}\right)r^2  + O(r^4), 
\end{eqnarray}
where $P_c$ is the central pressure of the star and $b_0>0$. In Appendix \ref{sec:expan} we show this perturbative solution up to terms of sixth order. This expansion in principle depends on four parameters: the theory parameter $\eta$, the central pressure $P_c$, the central density $\rho_c$, and the clock parameter of the scalar field $Q$. However, the central density and the central pressure will be related by the equation of state, as we will discuss below. The parameter $b_0$ can be absorbed by the standard rescaling of the time and radial coordinates. In addition the field equations only depend on the product $Q^2\eta$, so in practice we can restrict to the cases with $\eta=\pm 1$ and positive $Q$. This implies that
the solution has three physically relevant parameters: the central pressure (related to the total mass), $Q$ and the sign of $\eta$. 

However the perturbative solution is not regular for an arbitrary range of the physical parameters. For instance, regularity implies that, when $\eta>0$, a physical solution only exists when $\eta Q^2 < \frac{4 \kappa}{3}b_0$. Even more, if this condition holds then the scalar field is real close to the origin.{\color{black}{ In the $\eta<0$ case the solutions are always regular.}} However, the scalar field becomes imaginary close to the origin if 
$\abs{\eta}Q^2>b_0\frac{12 P_c \kappa}{ \left(2 \rho_c-3P_c \right)}$ and it is no longer clear if the solutions are physically relevant \cite{scalar,ludovic}.

The interior solution has to be matched with the exterior solution at the border of the star, which is found at $P(r=R_*)=0$. 

In addition to these regularity and junction conditions, we have to supplement the system of equations with an equation of state relating the energy density with the pressure, $\rho=\rho(p)$. We employ different models for the matter. A first approximation can be obtained using a relativistic polytrope:
\begin{equation}
\rho = \frac{K}{\Gamma+1} n^{\Gamma}+n, \ \ \ P=Kn^{\Gamma},
\end{equation}
with $n$ the baryon density, $K=1186$ and $\Gamma=2.34$. This simple model has been used previously in different settings \cite{Blazquez-Salcedo:2015ets,Motahar:2017blm}. 

\begin{figure} 
\includegraphics[width=8.5cm]{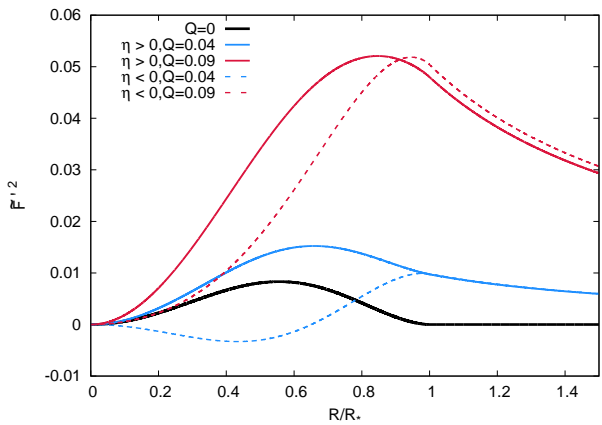}\\
\caption{Plot of $(F')^2$ as a function of $r/R_*$ for the polytropic EOS with $P_c = 2.5 \times 10^{34} \frac{\text{dyn}}{\text{cm}^2}$ and different values of $Q$ and $\eta$. 
}
\label{pic:stat2}
\end{figure}
\begin{figure} 
\includegraphics[width=8.5cm]{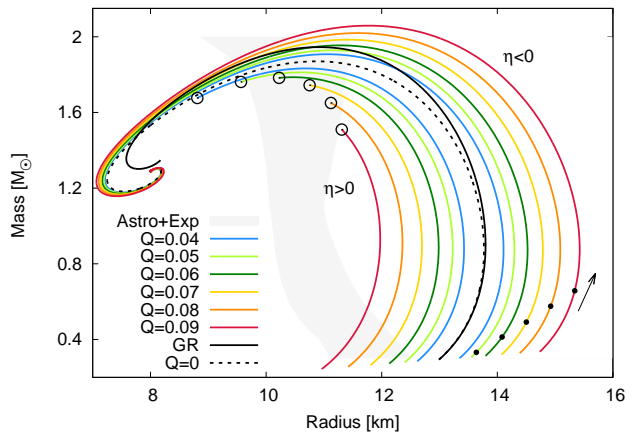}\\
\caption{Mass-radius relation for the polytropic equation of state. 
Open dots represent the points where a regular solution ceases to exist for $\eta>0$, and filled dots represent the points where the scalar field becomes imaginary for $\eta<0$ (see the discussion in the text). The arrow points in the direction of increasing pressure and an imaginary scalar field. The shaded area corresponds to the region favored by experiments and observations \cite{neutron_data}.}
\label{pic:stat1}
\end{figure}

We also consider realistic EOS which we implement using the piecewise polytropic interpolation presented in \cite{para_eos}. We choose a group of EOS with different matter content, providing a maximum mass of at least $\sim 2 M_{\odot}$, an observational constraint from pulsars PSR J1614-2230 \cite{Demorest:2010bx} and PSR J0348+0432 \cite{Antoniadis:2013pzd}. Following the nomenclature introduced in \cite{para_eos}, we choose SLy, APR4, WFF1 and WFF2 with plain nuclear matter; GNH3 and H4 for EOS containing hyperons; ALF2 and ALF4 for EOS containing quark matter.

We solve the set of Ordinary differential equations (\ref{tov1}) and (\ref{tov4}) numerically with a fourth order Runge-Kutta method \cite{rk41,rk42}, where we integrate the solution in the interior of the star, and match it at the border of the star at $r=R_*$ with the exterior solution. Typical profiles for the $(F')^2$ function are presented in Figure 
\ref{pic:stat2} as a function of $r/R_*$, where we use the polytropic EOS with $P_c = 2.5 \times 10^{34} \frac{\text{dyn}}{\text{cm}^2}$ and different values of $Q$ and $\eta$. Note that for this particular example with negative $\eta$ and $Q=0.04$, $(F')^2$ becomes negative in a region inside the star, hence the scalar field becomes imaginary. For $Q=0$ the scalar field vanishes outside the star, but it does not vanish in the interior, where it is real, and modifies the properties of the configuration with respect to the GR case. 

In Figure \ref{pic:stat1} we present the mass-radius diagram for the polytropic equation of state. With a black solid curve we mark the standard GR results, and with a dotted black curve the $Q=0$ results, which do not reduce to the GR case, due to the existence of a nontrivial scalar field hidden inside the configuration. The stars with $Q=0$ are in fact always regular and the scalar field is real everywhere inside the star. 

The colored curves below $Q=0$ correspond to the stars with $\eta>0$ and several values of $Q$. The curves stop at some critical points, marked with an open circle in the figure. Here the solutions stop being regular at the center, as discussed before. 

Above $Q=0$ we find the colored curves corresponding to $\eta<0$. Here the solutions are always regular, but they present an imaginary scalar field inside the star above the black dots, as predicted from the perturbative solution (\ref{pert_sol1}). The arrow points in the direction of increasing pressure. The shaded area corresponds to the region favored by experiments and observations \cite{neutron_data}. In Figure \ref{pic:MRfull} of Appendix \ref{sec:MR} we present similar figures for the mass-radius relation {\color{black}{from the other considered EOS}}. 

In general the $Q=0$ case presents slightly modified mass-radius relations {\color{black}{compared to}} GR, with a smaller value of the maximum mass and smaller values of the radius. The difference with respect to GR is, however, not very big (the maximum mass is between $4\%$ and $8\%$ less than in GR). Increasing $Q$ with $\eta>0$ reduces the maximum mass even more. For large enough values of $Q$, the curves are eventually too far below the $2 M_{\odot}$ limit, and the models are no longer physically relevant. Also note that the configurations tend to have smaller radii than in GR. With negative $\eta$ the opposite happens, and the maximum mass increases with $Q$ while the radius grows. However, the configurations possess imaginary scalar fields inside the star and the physical interpretation and relevance of these solutions are not clear.

\section{Axial perturbations}
\label{sec:pert}

Let us now focus on the static configurations ($Q=0$). As we have seen, outside the neutron star the scalar field vanishes and the metric coincides with the Schwarzschild solution. However, the scalar field is present inside the star, and modifies the way the matter content curves the space-time. Although the mass-radius relation differs very little from GR, it is interesting to know whether these scalarized stars possess similar resonant frequencies during the ringdown phase. We therefore study axial perturbations for these background solutions in order to extract the quasinormal modes describing the ringdown phase of the GWs and compare with the GR spectrum. 

Following the standard procedure \cite{Kokkotas:1999bd,Nollert:1999ji,Ferrari:2007dd}, we can always perturb the metric Ansatz (\ref{metric}) and scalar field (\ref{rest_field}) with a nonradial perturbation conveniently decomposed into tensorial spherical harmonics \cite{ten_sphe}. We hence focus on axial perturbations that do not affect the scalar field, pressure or density. If we treat the above solutions as the background (${g}^{(0)}_{\mu \nu}, {u}^{(0)}_{\mu}$), then we have to perturb the metric and the four velocity: 
\begin{eqnarray}
g_{\mu \nu} = {g}^{(0)}_{\mu \nu} + \epsilon h_{\mu \nu}^{(\text{axial})}\\
u_{\mu} = {u}^{(0)}_{\mu} + \epsilon \delta u_{\mu}^{(\text{axial})},
\end{eqnarray}
where we introduce the bookkeeping perturbation parameter $\epsilon$. 
In addition we perform a Laplace transformation to get rid of the time dependence, introducing the eigenvalue $\omega$ that will be in general a complex number whose real and imaginary parts determine the frequency ($\nu$) and the damping time ($\tau$) respectively, $\omega=\frac{\nu}{2\pi} + i\frac{1}{\tau}$. 

Hence an Ansatz for the axial perturbations in an appropriate gauge can be written as
\begin{widetext}
\begin{eqnarray}
h_{\mu\nu}^{(\text{axial})} &=& \int d\omega \, e^{-i\omega t}  \sum\limits_{l,m}
\left[
\begin{array}{c c c c}
	0 & 0 & -h_{0}	\frac{1}{\sin\theta}\frac{\partial}{\partial\varphi}Y_{lm} & h_{0}	\sin\theta\frac{\partial}{\partial\theta}Y_{lm} \\
	0 & 0 & -h_{1}	\frac{1}{\sin\theta}\frac{\partial}{\partial\varphi}Y_{lm}  
	& h_{1}	\sin\theta\frac{\partial}{\partial\theta}Y_{lm} \\
-h_{0}	\frac{1}{\sin\theta}\frac{\partial}{\partial\varphi}Y_{lm} & 
	-h_{1}	\frac{1}{\sin\theta}\frac{\partial}{\partial\varphi}Y_{lm} 
	& 0  & 0
 \\
h_{0}	\sin\theta\frac{\partial}{\partial\theta}Y_{lm} & h_{1}	\sin\theta\frac{\partial}{\partial\theta}Y_{lm}  
   & 0 & 0	
\end{array}
\right], \nonumber \\
\delta u^{(\text{axial}) \mu} &=& \int d\omega e^{- i \omega t} \sum_{\ell ,m} \frac{w_2}{\sin^2(\theta)} \left[ 0, 0, \frac{\partial}{\partial \theta} Y_{\ell m}, -\sin(\theta) \frac{\partial}{\partial \varphi} Y_{\ell m} \right] .
\end{eqnarray}
\end{widetext}

Introducing this Ansatz for the perturbations into the field equations (\ref{eq:field_eq}), and expanding up to first order in perturbation theory, we obtain the following set of relations:
{\color{black}{
\begin{eqnarray}
\label{eq_pert}
\frac{d {h_0}}{dr} &=& \frac{2}{r}{h_0} + \frac{i(2bn-\omega^2 r^2)}{\omega  r^2}{h_1}, \nonumber \\
\frac{d {h_1}}{dr} &=& -\frac{i \omega}{bf}\frac{4 \kappa - r^2 P}{4 \kappa + r^2 P}{h_0} 
\\ 
&+&\frac{4\kappa (f-1)- \left( 4f + 1\right)P r^2  - f \rho r^2 }{(r^2P+ 4 \kappa)rf}{h_1}, \nonumber \\ 
w_2 &=& \frac{{h_0}(P+\rho)}{r^2}, \nonumber
\end{eqnarray}
}}
where we have used the abbreviation $2n=(\ell-2)(\ell+1)$.
Note that this differs from GR inside the star, where we have the same relations for $w_2$ and $\frac{d h_0}{dr}$, but
\begin{eqnarray}
\frac{d h_1}{dr} = - \frac{i \omega}{bf} h_0  + \frac{4 \kappa(f-1)+(\rho -P)r^2 }{4\kappa r f} h_1 .
\end{eqnarray}
We see that
there is no continuous limit to GR unless $\rho=P=0$. 

Let us comment here that outside the star, since these equations then coincide with the ones from GR, it is possible to rewrite them as a second order differential equation (the Regge-Wheeler equation \cite{regge}). However inside the star
the resulting second order equation is different from GR. Its coefficients depend on the background configuration in a complicated way, and its comparison with the GR case is not enlightening. Hence for the calculation of the spectrum of QNMs we integrate the system of first order equations (\ref{eq_pert}), which possesses a simpler dependence on the background solution.

Since we are interested in purely outgoing gravitational waves that are radiated from the object, we should require the solutions of equation (\ref{eq_pert}) to behave asymptotically far from the star ($r \to \infty$) as

\begin{eqnarray}
h_0 = e^{-i \omega r_{\star}} \hat{H_0} \left(r + \hat{H_0} + \frac{\hat{H_1}}{r}+ O(r^{-2}) \right), \\
h_1 = e^{-i \omega r_{\star}} \hat{H_0} \left(r + \hat{H_2} + \frac{\hat{H_3}}{r}+ O(r^{-2}) \right),
\label{infinity_pert}
\end{eqnarray}
 where $\hat{H_0}$ is an arbitrary amplitude and $\frac{dr_{\star}}{dr} =  \frac{1}{\sqrt{b f}} $. The constants are determined by the perturbation equations (\ref{eq_pert}), and satisfy
 \begin{eqnarray}
&& \hat{H_1}= \frac{n}{i\omega}, \
 \hat{H_2}= \frac{M}{2i\omega}-\frac{n(n+1)}{2\omega^2}, \
 \hat{H_3}= 2M + \frac{n+1}{i\omega}, \nonumber \\
&& \hat{H_4}= 4M^2 +\frac{2M}{i\omega}\left(n+\frac{1}{4} \right) - \frac{n(n+1)}{2\omega^2},
 \end{eqnarray}
where $M$ is the mass of the Schwarzschild solution outside the star.

Regularity of the perturbation at the center of the star implies that the perturbation behaves as
\begin{eqnarray}
\label{origin_pert}
&& h_0 = H_0 r^\ell \left( 1 + \frac{H_1}{2\ell+1} r^2 + O(r^4) \right), \\
&& h_1 = \frac{- i \omega}{b_0(\ell+1)} H_0 r^{\ell+1} \left( 1 + \frac{H_2}{2\ell+1} r^2 + O(r^4) \right), \nonumber 
\end{eqnarray}
with the amplitude $H_0$ being an arbitrary constant and
\begin{eqnarray}
\label{origin_pert_aux}
&& H_1 = \frac{\ell-2}{8\kappa}\left[(\ell-4)P_c + \ell \rho_c \right]-\frac{\omega^2 (\ell+3)}{2 b_0 (\ell+1)},  \\
&& H_2 = \frac{-\omega^2}{2b_0}+ 
%\\  &&
 \frac{3(\ell^2-8\ell-2)P_c + (3\ell^2-4\ell-2)\rho_c}{24\kappa}
. \nonumber
\end{eqnarray}

In order to obtain the QNMs, we use a shooting method. Once a background solution is obtained using the procedure described in the previous section, we divide the solution into two  different domains. The first domain goes from the origin $r=0$ to the border of the star $r=R_*$. The second domain goes from the border of the star up to a relatively large value of the radial coordinate, $r=R_{i}\gg R_*$, a parameter that we can adjust in order to optimize the procedure. We solve the perturbation equations (\ref{eq_pert}) in the first domain imposing the regularity condition (\ref{origin_pert}) at $r=0$. In the second domain we solve the perturbation equations (\ref{eq_pert}) imposing the outgoing wave behavior (\ref{infinity_pert}) at $r=R_{i}$. The solution of the perturbation equations is obtained using the package \textsc{Colsys} \cite{colsys} over a cubic Hermite spline-interpolated background solution obtained from the Runge-Kutta method \cite{rk41,rk42}, with a mesh of $1000$ points inside the star. The resonance is obtained when the perturbations of both domains are continuous across the border of the star \cite{detweiler,osci_star}, which only happens at discrete values of $\omega$. In order to obtain the modes we implement a search algorithm based on the gradient descent \cite{borwein,grad}.

\section{Spectrum and universal relations}
\label{results}

Following this procedure it is possible to generate the spectrum of the $Q=0$ neutron stars. We hence focus the discussion on the $\ell=2$ fundamental curvature modes, which are typically the ones expected to be strongly excited during the merger process.

\begin{figure} 
\begin{center}
\includegraphics[width=9.5cm]{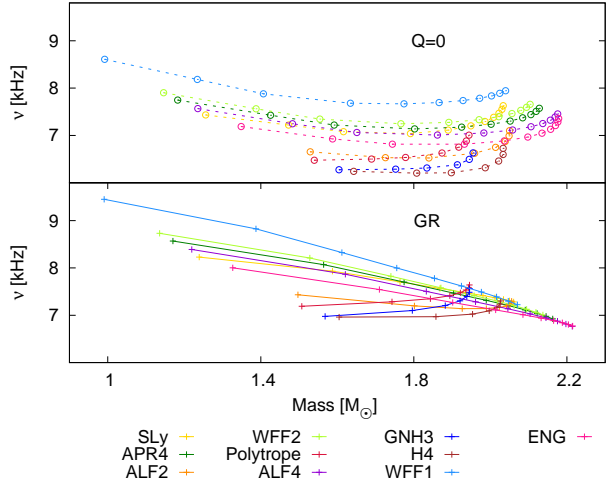}\\
\caption{Frequency in $kHz$ versus the total mass in solar masses for the fundamental $\ell=2$ mode. In the upper panel we show the $Q=0$ scalarized stars with dashed lines, and in the lower panel the GR stars with solid lines. Each color represents a different equation of state.}
\label{pic:plot5}
\end{center}
\end{figure}

In Figure \ref{pic:plot5} we show the frequency as a function of the total mass of the star. In the upper panel we present the frequencies for the $Q=0$ scalarized stars, while the lower panel presents the same results for GR. Each EOS is represented with a different color: SLy in yellow, WFF2 in light green, GNH3 in dark blue, ENG in pink, APR4 in dark green, the polytrope in red, H4 in brown, ALF2 in orange, ALF4 in purple, and WFF1 in light blue. A comparison between the two theories reveals that the effect of the scalar field on the frequency is not so important as the effect of changing the equation of state, which resembles what happens for the mass-radius relations. From this figure it is not possible to extract a general behavior, since the introduction of the scalar field can either increase or decrease (slightly) the value of the frequency {\color{black}{compared to a}} GR configuration, depending on the particular EOS and value of the mass.

\begin{figure} 
\begin{center}
\includegraphics[width=9cm]{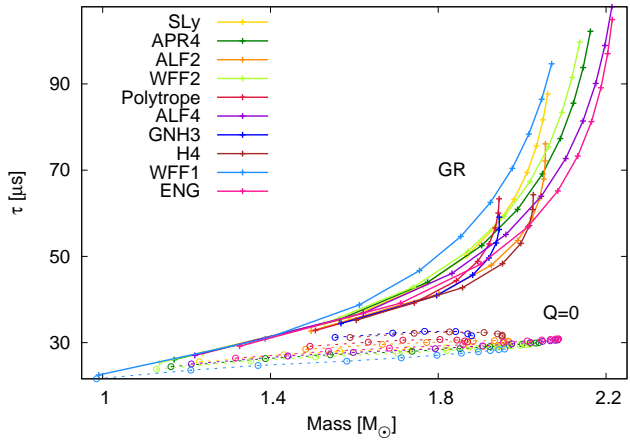}\\
\caption{Damping time in $\mu s$ versus the total mass in solar masses for the fundamental $\ell=2$ mode. The solid lines represent GR neutron stars, and the dashed lines the $Q=0$ scalarized stars. Each color represents a different equation of state.}
\label{pic:plot5_2}
\end{center}
\end{figure}

In Figure \ref{pic:plot5_2}, we show the damping time as a function of the total mass. Here the solid lines represent GR, and the dashed lines the $Q=0$ scalarized configurations. We can see that for the low mass stars (around $1 M_{\odot}$) both theories predict almost the same damping times. However, close to the maximum mass of each EOS, the damping times predicted in each theory are very different: the $Q=0$ scalarized configurations possess much shorter damping times than the GR stars. While the ringdown phase in GR can exist for $60-100 \mu s$ (depending on the EOS), the $Q=0$ configurations have a decay time of $\sim 30 \mu s$, meaning the signal would be damped almost 2-3 times faster than in GR. Interestingly the damping times of these configurations does not change much with the mass or the EOS. 

Hence a first result is that, although we do not observe a significant deviation in the frequency part, the damping times of the most massive scalarized stars are much shorter than the corresponding damping times in GR.

\begin{figure} 
\begin{center}
\includegraphics[width=9cm]{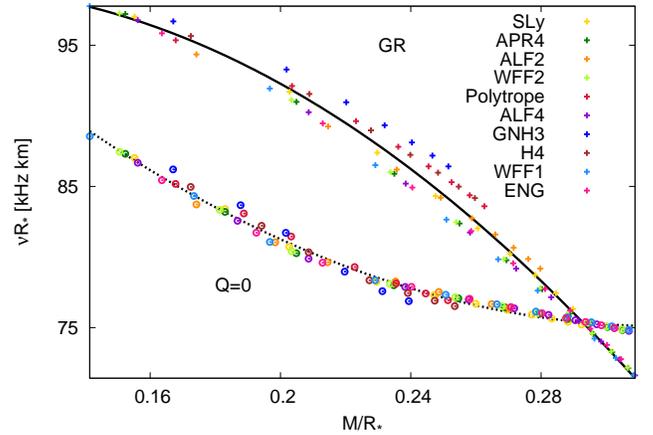}\\
\caption{The frequency scaled to the radius ($\nu R_*$) versus the compactness of the star. Crosses represent the modes of GR stars, and open circles represent the ones of $Q=0$ scalarized stars. The solid black line shows the phenomenological relation (\ref{ph_rel1}) for GR modes, while the dashed black curve shows the relation (\ref{ph_rel2_1}) for $Q=0$ scalarized stars.}
\label{pic:plot7}
\end{center}
\end{figure}

\begin{figure} 
\begin{center}
\includegraphics[width=9cm]{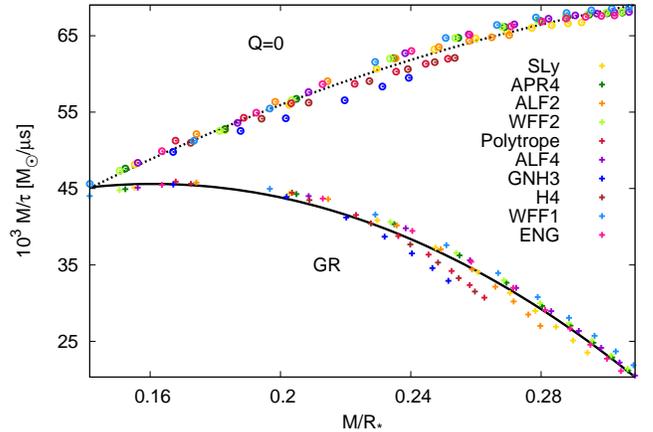}\\
\caption{Similar to Figure \ref{pic:plot7}, but showing the inverse of the damping time scaled with the mass ($\frac{10^3 M}{\tau}$) versus the compactness of the neutron stars. }
\label{pic:plot7_2}
\end{center}
\end{figure}

However, we can also observe that the indetermination in the matter model is large enough to obscure the effect of changing the gravitational theory. To avoid this, let us now study the standard matter independent phenomenological relations that are known to approximately hold for neutron stars in GR. Such relations have been studied before in \cite{Kokkotas:1999mn,Benhar:2004xg,BlazquezSalcedo:2012pd,Blazquez-Salcedo:2013jka} in the context of GR, and in dilatonic-Einstein-Gauss-Bonnet theory in \cite{Blazquez-Salcedo:2015ets}. 

Concerning the frequency a universal relation can be found if one considers the scaled quantity $\nu R_*$ as a function of the compactness of the star $M/R_*${\color{black}{, which is shown in Figure \ref{pic:plot7}.}} The square points mark the GR modes, while the open circles mark the modes of $Q=0$ scalarized stars. Each color represents a different EOS, and we follow the convention of Figure \ref{pic:plot5}. The figure reveals that, although in each theory the relation is quite independent of the matter content, the universal relation is completely different in each theory. 
In the GR case the relation can be approximately described by a quadratic function of the compactness\newpage

\begin{eqnarray}
\label{ph_rel1}
&& \nu_{GR}(\text{kHz})=\frac{1}{R_*(\text{km})} \left[\left( -589.20 \pm 45.44 \right) \left(\frac{M}{R_*}  \right)^2 \right. \nonumber \\ && \ \ \left. + \left( 108.15 \pm 21.34 \right) \left(\frac{M}{R_*}  \right) + \left( 94.21 \pm 2.4 \right)   \right],
\end{eqnarray} 
which is compatible with the results from \cite{Kokkotas:1999mn,Benhar:2004xg,BlazquezSalcedo:2012pd,Blazquez-Salcedo:2013jka}. We mark this relation in Figure \ref{pic:plot7} with a solid black line. For the $Q=0$ scalarized stars, however, the relation is also quadratic but given by 
\begin{eqnarray}
\label{ph_rel2_1}
&& \nu_{}(\text{kHz}) = \frac{1}{R_*(\text{km})} \left[\left( 444.87 \pm 18.26 \right) \left(\frac{M}{R_*}  \right)^2 \right. \nonumber \\  && \ \ \left. + \left( -282.66 \pm 8.54 \right) \left(\frac{M}{R_*}  \right) + \left( 120.00 \pm 0.97 \right)   \right].
\end{eqnarray}
We mark this relation in Figure \ref{pic:plot7} with a black dashed line. The dashed curve is well below the solid curve in the low  compactness region, while both curves tend to get closer for higher compactness, eventually crossing each other around $M/R_*=0.29$.

Concerning the damping time, a nice phenomenological relation quite independent of the matter model can be found for $\frac{10^3 M}{\tau}$ as a function of the compactness $M/R_*$. In Figure \ref{pic:plot7_2} we show the damping times scaled in this way. Again the square points mark the GR modes and the open circles the modes of the $Q=0$ case. Here we can clearly see a huge difference in the universal relations between the two theories. In GR the scaled frequency tends to decrease, when the compactness is increased, while for the $Q=0$ scalarized stars the behavior is the opposite, and the frequency increases with the compactness. Highly compact stars ($M/R_*\sim 0.3$) with a nontrivial scalar field inside present a huge deviation in this scaled quantity with respect to the GR case. 

Again, the phenomenological relation in the GR case can be approximately described by a quadratic function of the compactness
\begin{eqnarray}
\label{ph_rel1_2}
&& \frac{10^3}{\tau_{GR}(\mu\text{s})} = \frac{1}{M(\text{M}_{\odot})} \left[\left( -1155.7 \pm 57.54 \right) \left(\frac{M}{R_*}  \right)^2 \right. \nonumber \\ && \ \ \left. + \left( 371.97 \pm 27.02 \right) \left(\frac{M}{R_*}  \right) + \left( 15.67 \pm 3.10 \right)   \right],
\end{eqnarray}
which is again compatible with \cite{Kokkotas:1999mn,Benhar:2004xg,BlazquezSalcedo:2012pd,Blazquez-Salcedo:2013jka}. In Figure \ref{pic:plot7_2} this is shown as a solid black line.
The $Q=0$ scalarized stars approximately satisfy
\begin{eqnarray}
\label{ph_rel2}
&& \frac{10^3}{\tau_{}(\mu\text{s})} = \frac{1}{M(\text{M}_{\odot})} \left[\left( -391.62 \pm 45.04 \right) \left(\frac{M}{R_*}  \right)^2 \right. \nonumber \\ && \ \ \left. + \left( 319.48 \pm 21.06 \right) \left(\frac{M}{R_*}  \right) + \left( 7.71 \pm 2.39 \right)   \right] ,
\end{eqnarray}
and are represented with a black dashed line.
\begin{figure} 
\begin{center}
\includegraphics[width=9cm]{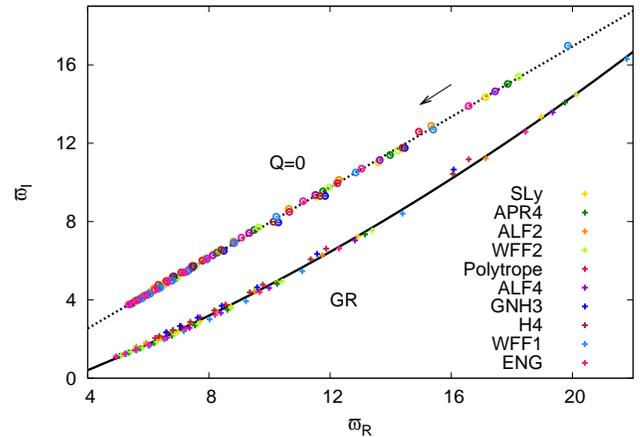}\\
\caption{$\varpi_I$ versus $\varpi_R$ for the $\ell=2$ fundamental modes of neutron stars with different equations of state. Crosses and circles represent the modes of GR and $Q=0$ stars, respectively. The black solid line shows the universal relation (\ref{eq:ph1}) for GR, and the black dashed line the universal relation (\ref{eq:ph2}) for $Q=0$ scalarized stars.
The arrow marks the direction of increasing central pressure and compactness.}
\label{pic:plot8}
\end{center}
\end{figure}
\begin{figure} 
\begin{center}
\includegraphics[width=9cm]{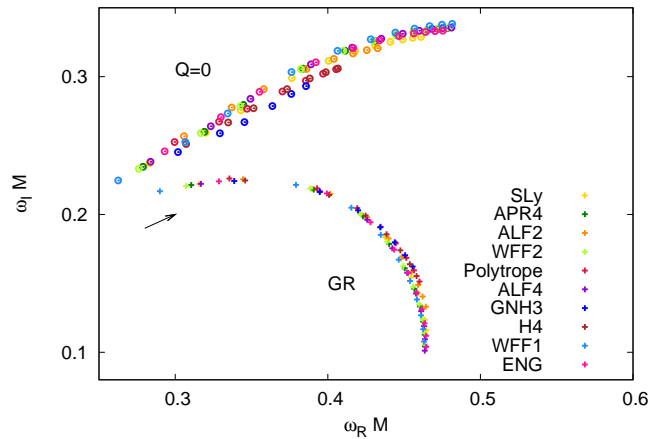}\\
\caption{Similar to Figure \ref{pic:plot8}, but showing the dimensionless quantities $\omega_I M$ versus $\omega_R M$.}
\label{pic:plot8_2}
\end{center}
\end{figure}

An alternative phenomenological relation can be obtained by scaling the eigenvalue $\omega$ with the central pressure of the star \cite{BlazquezSalcedo:2012pd,Blazquez-Salcedo:2013jka,Blazquez-Salcedo:2015ets}. If we define the dimensionless quantities  $\varpi_R = \frac{\omega_{\Re}}{\sqrt{P_c}}$ and  $\varpi_I = \frac{\omega_{\Im}}{\sqrt{P_c}}$, then it is possible to see that in GR, the modes follow a universal quadratic relation of the form
\begin{eqnarray}
\label{eq:ph1}
&& \varpi_{I_{GR}} =\left( 0.0148 \pm 0.0008 \right) \varpi_{R_{GR}}^2 \nonumber \\ && \ \ + \left(0.519 \pm 0.020 \right) \varpi_{R_{GR}} + \left( -1.90 \pm 0.10 \right)   ,
\end{eqnarray}
which is compatible with \cite{BlazquezSalcedo:2012pd,Blazquez-Salcedo:2013jka,Blazquez-Salcedo:2015ets} within the precision of the numerical procedure.
Interestingly, for the $Q=0$ scalarized stars, a similar universal behavior can be obtained, but this time the relation is best described by a linear relation of the type
\begin{eqnarray}
\label{eq:ph2}
&& \varpi_{I_{}} =  \left(0.900 \pm 0.002 \right) \varpi_{R_{}} + \left( -1.06 \pm 0.02 \right) .
\end{eqnarray}
We present these scaled relations in Figure \ref{pic:plot8}, where we show $\varpi_I$ as a function of $\varpi_R$. With a black solid line we show the GR relation (\ref{eq:ph1}), and with a black dashed line the relation (\ref{eq:ph2}).

Finally, we show a similar matter independent relation, this time by scaling the frequency and the damping time only with the total mass of the neutron star. In Figure \ref{pic:plot8_2} we show the analogous diagram to Figure \ref{pic:plot8}, but now the dimensionless parameter $\omega_I M$ as a function of $\omega_R M$. This Figure also shows an important difference between the spectrum of stars in GR and in the $Q=0$ case, since the scaled imaginary part of the frequency can be very different between the theories, as we have already shown in Figure \ref{pic:plot7_2}.

\section{Conclusions and outlook}
\label{sec:con}
In this paper we have presented the axial quasinormal modes of scalarized neutron stars in a restricted sector of Horndeski gravity, known as the nonminimal derivative coupling sector. 

In section II we have briefly reviewed the solutions from \cite{scalar,ludovic}, where the metric is static, but the scalar field can in principle have a linear temporal dependence controlled by the ''clock'' parameter $Q$. These solutions are described by the Schwarzschild metric outside the star. In the special case that $Q=0$, the scalar field is time independent and vanishes outside the star. The resulting neutron stars have a hidden scalar field in their interior, which changes the way matter couples to gravity. In section III we have studied the axial perturbations of the $Q=0$ configurations.

Although the mass-radius relations of the $Q=0$ scalarized stars are quite close to the ones of GR (differing mostly by $8\%$ at the maximum mass), the analysis of the spectrum of quasinormal modes in section III, and in particular, the fundamental $\ell=2$ mode, reveals some important differences. The scalarized stars possess much shorter damping times in their ringdown phase. When compared with GR, the damping can be 2 or 3 times faster, depending on the EOS. Moreover, we have calculated several universal relations in terms of global parameters of the stars, and we have found that the relations remain matter independent, but they can deviate drastically from the GR case, in particular when a scaling with global quantities like the total mass and radius is used.

A similar analysis of the polar perturbations remains to be done. In this case, one should include the perturbations of the density and pressure, but also of the scalar field, inside and outside of the star. The existence of a non-trivial scalar field will produce a new family of scalar-led modes \cite{Blazquez-Salcedo:2016enn,Blazquez-Salcedo:2016yka,Blazquez-Salcedo:2017txk}, in addition to the standard space-time and fluid modes, resulting in a richer spectrum when compared with GR. In addition, it would be interesting to generalize this analysis to the case of neutron stars with a time dependent scalar field ($Q\neq 0$), at least in the region of parameter space, where these solutions could have some astrophysical relevance. 

\begin{figure*}
\includegraphics[width=17.8cm,angle=0]{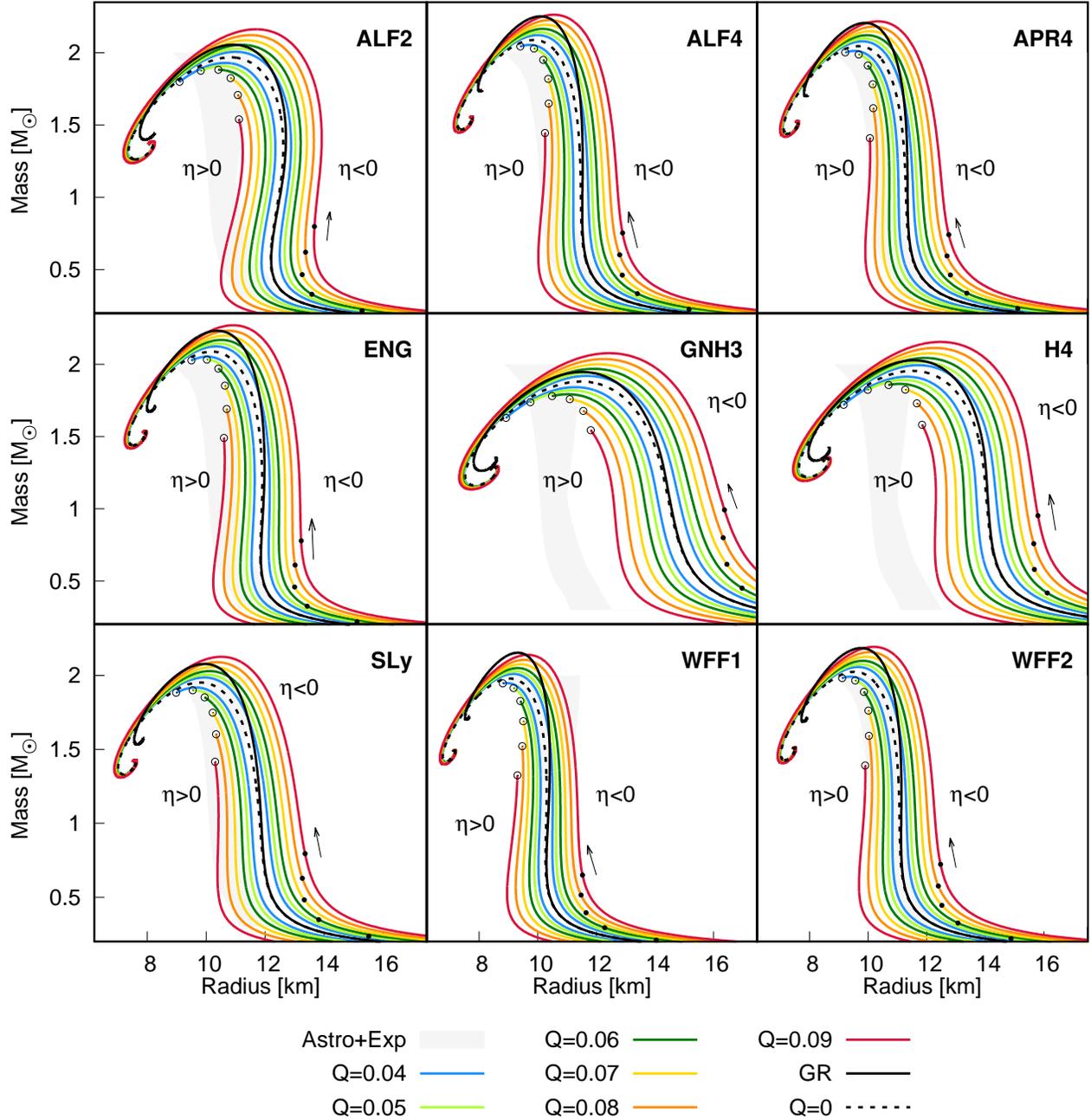}\\
\caption{Mass-radius relation for the nine considered realistic EOS, and different values of $Q$ and the sign of $\eta$.
}
\label{pic:MRfull}
\end{figure*}

\section*{ACKNOWLEDGEMENTS}

The authors want to thank Jutta Kunz for discussions.
We gratefully acknowledge support by the Deutsche Forschungsgemeinschaft (DFG), in particular, within the framework of the DFG Research Training group 1620 {\it Models of gravity}. J.L.B.S. acknowledges support from FP7, Marie Curie Actions, People, International Research Staff Exchange Scheme (IRSES- 606096).

\appendix

\section{Perturbative expansion around the center of the star}
\label{sec:expan}

In this section we present the perturbative solution describing a spherically symmetric neutron star close to the center ($r=0$). Up to sixth order, the Ansatz functions of equation (\ref{metric}) and (\ref{rest_field}) possess the following expansions:

\begin{eqnarray}
&f(r) \approx 1 + \hat{\mathfrak{f}}_2 r^2 + \hat{\mathfrak{f}}_4 r^4 + O(r^6), \\
&b(r) \approx b_0 + \hat{\mathfrak{b}}_2 r^2 + \hat{\mathfrak{b}}_4 r^4 + O(r^6), \nonumber \\
&P(r) \approx P_c + \hat{\mathfrak{P}}_2 r^2 + \hat{\mathfrak{P}}_4 r^4 + O(r^6), \nonumber \\
&\rho(r) \approx \rho_c + \hat{\rho_2} r^2 + \hat{\rho_4} r^4 + O(r^6), \nonumber \\
&F(r) \approx F_c + \hat{\mathfrak{F}}_2 r^2 + \hat{\mathfrak{F}}_4 r^4 + O(r^6). \nonumber 
\end{eqnarray}
The coefficients are determined by the modified TOV equations (\ref{tov1}) and the scalar field equation (\ref{tov4}). They are:
\begin{eqnarray}
&&\hat{\mathfrak{f}}_2 = \frac{2}{3} b_0 \frac{3 P_c + \rho_c}{\zeta},  \\
&&\hat{\mathfrak{b}}_2 = -\frac{1}{2} b_0 \hat{\mathfrak{f}}_2, \nonumber \\
&&\hat{\mathfrak{P}}_2 = \frac{1}{4} (P_c + \rho_c) \hat{\mathfrak{f}}_2, \nonumber \\
&&\hat{\mathfrak{F}}_2 = \sqrt{\frac{1}{4 b_0 \eta}(b_0 P_c-\hat{\mathfrak{f}}_2 Q^2\eta)} \nonumber \\
&&\hat{\mathfrak{f}}_4 = \frac{2}{5}\frac{b_0 \hat{\rho_2}}{\zeta} + \frac{2}{5} \frac{b_0^2 P_c}{\zeta^2}(3P_c+\rho_c)- \nonumber \nonumber \\
 && \ \ \ \frac{b_0 \hat{\mathfrak{f}}_2}{10\zeta^2}\big[(51 P_c + 3\rho_c)Q^2\eta +(4 P_c + 20\rho_c)b_0\kappa\big], \nonumber \\
&&\hat{\mathfrak{b}}_4 = \frac{1}{4}b_0(\hat{\mathfrak{f}}_2^2-\hat{\mathfrak{f}}_4)-\frac{1}{4}\hat{\mathfrak{b}}_2 \hat{\mathfrak{f}}_2, \nonumber \\
&&\hat{\mathfrak{P}}_4 = \frac{1}{8} \hat{\mathfrak{f}}_4 (P_c + \rho_c) + \frac{1}{8} \hat{\mathfrak{f}}_2 \hat{\rho_2} - \frac{3}{32} \hat{\mathfrak{f}}_2^2 (P_c + \rho_c), \nonumber \\
&&\hat{\mathfrak{F}}_4 = \frac{1}{16 \hat{\mathfrak{F}}_2 \eta b_0}\big[
(\frac{\hat{\mathfrak{f}}_2^2}{2}-\hat{\mathfrak{f}}_4)Q^2\eta-\frac{1}{4}(3P_c - \rho_c)\hat{\mathfrak{f}}_2 b_0
\big], \nonumber 
\end{eqnarray}

with $\zeta=3 Q^2\eta-4 b_0 \kappa$.

\

\section{Mass-radius relations for realistic equations of state}
\label{sec:MR}

In Figure \ref{pic:MRfull} we present the mass radius relations for the nine considered realistic EOS in this paper: SLy, APR4, WFF1, WFF2, GNH3, H4, ALF2 and ALF4, each one of them in a different panel. Each panel is similar to Figure \nolinebreak \ref{pic:stat1} for the polytropic EOS. The black solid lines are the GR results. The dotted black curves are the $Q=0$ configurations. The colored curves below the $Q=0$ dotted black curve correspond to the stars with $\eta>0$, while the colored curves above the $Q=0$ dotted black curve correspond to $\eta<0$. The open circles represent the end of the presence of regular solutions, while all solutions above the solid black dots possess an imaginary scalar field.

\bibliographystyle{unsrt}

\end{document}